\documentclass[aps,pre,twocolumn,final,10pt,groupedaddress,showpacs]{revtex4-1}

\usepackage[latin1]{inputenc}
\usepackage[english]{babel}
\usepackage{graphicx}
\usepackage{float}
\usepackage{psfrag}
\usepackage{amssymb}
\usepackage{amsmath}
\usepackage{amscd}
\usepackage{eucal}
\usepackage{color}
\usepackage{bm}
\newcommand{\Kf}{K\hspace{-0.8mm}f}
\input{epsf}
\begin{document}
\renewcommand{\figurename}{Figure}

\title{Global Robustness vs. Local Vulnerabilities in Complex Synchronous Networks
}
\author{Melvyn Tyloo\textsuperscript{1,3} and Philippe Jacquod\textsuperscript{2,3}}
\affiliation{\textsuperscript{1} Institute of Physics, \'Ecole Polytechnique F\'ed\'erale de Lausanne (EPFL), CH-1015 Lausanne, Switzerland. \\
\textsuperscript{2} Department of Quantum Matter Physics, University of Geneva, CH-1211 Geneva, Switzerland\\
\textsuperscript{3} School of Engineering, University of Applied Sciences of Western Switzerland HES-SO, CH-1951 Sion, Switzerland. }

\date{\today}

\begin{abstract}
In complex network-coupled dynamical systems, two questions of central importance are how to identify the most vulnerable components and how to devise a network making the overall system more robust to external perturbations. To address these two questions, we investigate the response of complex networks of coupled oscillators to local perturbations.
We quantify the magnitude of the resulting excursion away from the unperturbed synchronous state 
through quadratic performance measures in the angle or frequency deviations. 
We find that the most fragile oscillators in a given network are 
identified by centralities constructed from network resistance distances. Further defining the global robustness of the system from the average response over ensembles of homogeneously distributed perturbations, we find that it is given by a family of topological indices known as generalized Kirchhoff indices. Both resistance centralities and Kirchhoff indices are obtained from a spectral decomposition of the stability matrix of the unperturbed dynamics and can be expressed in terms of resistance distances. We investigate the properties of these topological indices in small-world and regular networks. In the case of oscillators with homogeneous inertia and damping coefficients, we find that inertia only has small effects on robustness of coupled oscillators. Numerical results illustrate the validity of the theory.
\end{abstract}

\maketitle
\section{Introduction}
Complex networks are widely used to model nature- as well as man-made coupled dynamical systems \cite{Rod16}. Physical realizations of such systems range from microscopic Josephson junction arrays \cite{Wie98} and interacting molecules in chemical reactions \cite{Kur84b,Kur75} to macroscopic  high voltage electric power grids \cite{Ber81} and communication or social networks \cite{Str17,Bar16}. Individual elements are represented by nodes in a complex network, which have internal parameters and degrees of freedom. The latter are governed by differential equations that depend on both the internal dynamics of the individual elements and the coupling to the adjacent nodes. Two central questions are (i) how to identify nodes, which, once attacked, perturbed or removed, have the most dramatic effect on the overall dynamics of the coupled system and (ii) how to devise a coupling network guaranteeing robustness of the system against random external perturbations. Attempts to answer such questions are often based on complex network theory, numerically relating dynamical effects to graph-theoretic metrics. This approach has been often criticized, e.g. in 
Ref.~\cite{Bor06,Bol14,Hin10}, because (i) it gives no {\it a priori} criterion for which metric should be considered in which situation and (ii) it does not directly indorporate the intrinsic dynamics of the network-coupled system.

 Here we propose an altogether different analytical approach. First, we use robustness performance measures that quantify the excursion during the transient dynamics following a perturbation. Second, we spectrally decompose the coupling matrix to calculate the response of the system to some external perturbations.
\begin{figure}[!h]
\begin{center}
\includegraphics[width=0.88\columnwidth]{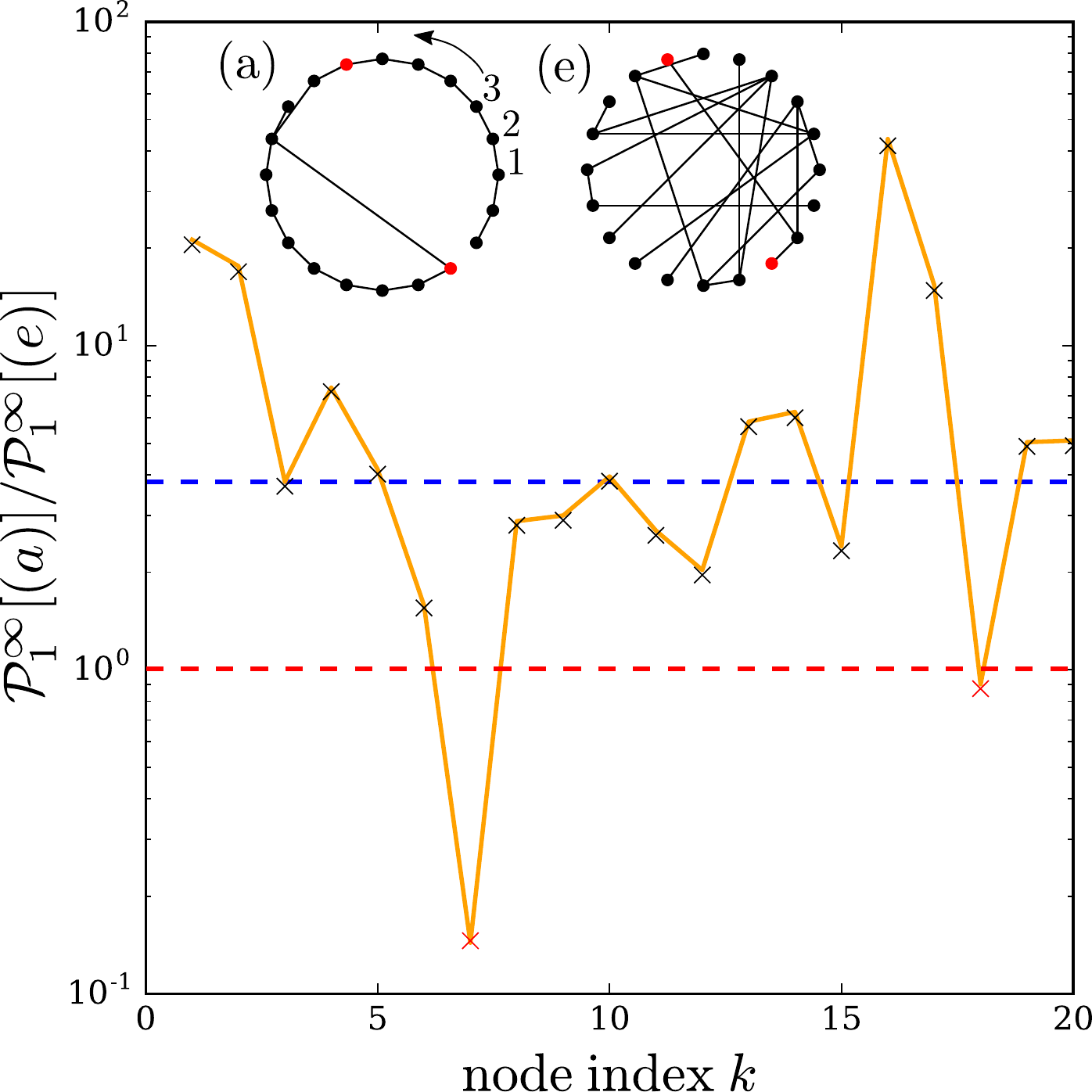}
\caption{Ratio of the performance measures $\mathcal{P}_1$ for graph (a) vs. (e) of Fig.~\ref{fig1} (shown in the insets), for a quench perturbation of magnitude $\delta P_0=0.01$ and duration $\gamma\tau_0=500$ on node $k$ (see text). On average, graph (e) is four times more robust to external perturbations than graph (a) (blue dashed line). However, some nodes of graph (a) can be more robust than those of graph (e) (red crosses correspond to quench perturbation applied on the red nodes shown in the inset). Both specific local vulnerabilities (crosses) and global averaged robustness (blue dashed lines) are well predicted by combinations of local centralities, and global topological indices
(orange solid line, see text).}\label{fig1_intro}
\end{center}
\end{figure}
Third, by direct calculation, we  relate the obtained analytical expressions for performance measures (i)  to local centralities when analyzing local vulnerabilities, and (ii) to global topological indices when assessing global robustness of the networked system. Following these steps, we identify a new class of local and global topological indices that characterize robustness of synchrony of complex network-coupled oscillators. Our method builds up on investigations of 
consensus algorithms \cite{Bam12,Gru18}, electric power systems \cite{Poo17,Pag17,Sia16} and coupled oscillators systems \cite{Tyl18a,Tyl18b}. Already implicitly present in Refs.~\cite{Gru18,Pag17,Sia16}, the Kirchhoff index was first identified as a global robustness quantifier in our earlier work, Ref.~\cite{Tyl18a}. Local vulnerabilities have been more recently connected to centralities related to the resistance distance \cite{Gru18,Tyl18b}.

In this manuscript, we investigate vulnerabilities and global robustness of synchronous 
network-coupled oscillators. Frequency synchronization often occurs in such systems when the 
coupling between individual oscillators is strong enough that they start to oscillate at the same
frequency, even when their natural frequency is not homogeneous~\cite{Str04,Pik03}. 
Frequency synchronization has attracted a large interest, in particular, the
robustness of the synchronous state has been studied from a variety of points of view. 
One may for instance consider the linear stability of the synchronous state~\cite{Pec98}, the range of
network parameters where synchrony occurs~\cite{Bar02,Cha05,Zho06}, the 
volume of the basin of attraction of the synchronous state~\cite{Wil06,Men13,Del17b}, the influence of noise
on the synchronous state, in particular how it can lead to desynchronization 
or drive the system to another synchronous state \cite{Dev12,Hin16,Scha17,Hin18,Tyl18c,Hin19}, how disturbances spread across the network \cite{Ket16,Wol18,Pag19}, or even how topological changes affect synchrony~\cite{Col16b,Sol17,Col17b}. Here, we investigate the
robustness of the synchronous state against external perturbations. For both local and ensemble-averaged perturbations on oscillators with identical dynamical parameters,
we find that the robustness 
 of the synchronous state is given by a new family of topological indices based on 
 the resistance distance \cite{Kle93,Tyl18a,Tyl18b}. This is illustrated in Fig.~\ref{fig1_intro} which shows the ratio between performance measures defined in Eq.~\eqref{eq:c1}, numerically obtained by perturbing each node of graphs (a) and (e) shown in the inset. Even if the average value of the performance measure is lower for graph (e), the latter can be more strongly sensitive to certain
local perturbations. Below we show that specific local vulnerabilities and global averaged robustness  are determined by nodal centralities and global topological indices (orange). 

The manuscript is organized as follows. Section \ref{section2} recalls the definition of the resistance distance and generalizes it to graphs corresponding to powers of the Laplacian matrix. Section \ref{section3} describes our model of coupled oscillators, briefly discusses synchronized states
and evaluates how they respond to external perturbations. 
Performance measures quantifying this response are also introduced and calculated for quench perturbations. 
Sections \ref{section4} numerically illustrates the theory on different graphs for local and global vulnerabilities. An analysis of Kirchhoff indices in both small-world and regular networks is also done. We conclude in Section \ref{section7}.

\section{Resistance Distances, Centralities and Kirchhoff Indices}\label{section2}
The resistance distance $\Omega_{ij}^{(1)}$ is a graph-theoretic metric with an intuitive physical interpretation \cite{Kle93}. To any graph, one associates an electrical network of resistors whose capacities are given by the inverse of the edge weights. In this case, $\Omega_{ij}^{(1)}$ is the effective resistance between $i$ and $j$, i.e. the voltage that develops between $i$ and $j$ when a unit current is injected at $i$ and collected at $j$ with no injection nor collection at any other node. The resistance distance can be expressed with the network Laplacian matrix $\mathbb{L}$ as
\begin{eqnarray}\label{eq:rdist}
\Omega_{ij}^{(1)}=\mathbb{L}^\dagger_{ii} + \mathbb{L}^\dagger_{jj} - \mathbb{L}^\dagger_{ij} - \mathbb{L}^\dagger_{ji} \; ,
\end{eqnarray}
where $\mathbb{L}^\dagger$ is the Moore-Penrose pseudo inverse of $\mathbb{L}$.
The resistance distance can be formulated in a convenient way using eigenvectors ${\bf u}_\alpha$ and eigenvalues $\lambda_\alpha$ of $\mathbb{L}$. It is given by \cite{Xia03,Col17b},
\begin{eqnarray}\label{eq:Omega_eig}
\Omega_{ij}^{(1)}=\sum_{\alpha \ge 2}\frac{(u_{\alpha,i}-u_{\alpha,j})^2}{\lambda_\alpha} \; ,
\end{eqnarray} 
where the zero-eigenvector of $\mathbb{L}$ corresponding to $\lambda_1=0$ is omitted in the sum. The resistance distance is a graph-theoretic distance 
metric because (i) $\Omega_{ii}^{(1)}=0$, $\forall i$, (ii) $\Omega_{ij}\ge 0$ $\forall i,j$,  and (iii) $\Omega_{ij}^{(1)}+\Omega_{jk}^{(1)}\ge \Omega_{ik}^{(1)}$, $\forall i,j,k$ (triangle inequality) \cite{Kle93}. 

A measure of nodal centrality
 is given by the inverse of the average resistance distance from any node $k$
to all other network nodes,
\begin{align}\label{eq:C1}
\begin{split}
C_1(k)=\left[ n^{-1}\sum_j \Omega_{kj}^{(1)} \right]^{-1}=\left[ \sum_{\alpha\ge 2}\frac{u_{\alpha,k}^2}{\lambda_\alpha} + n^{-2}\Kf_1 \right]^{-1}\; .
\end{split}
\end{align}
It is a closeness centrality in the usual sense~\cite{Bol14}, meaning in particular that  
large values of $C_1(k)$ indicate nodes $k$ that are central in the network
according to the resistance distance $\Omega^{(1)}_{ij}$. The second term in bracket on the
right-hand-side of  Eq.~(\ref{eq:C1}) is a graph topological index known as the Kirchhoff index of the network and defined by \cite{Kle93},
\begin{eqnarray}\label{eq:Kirchhoff}
\Kf_1&=&\sum_{i<j}\Omega_{ij}^{(1)}=n\sum_{\alpha\ge 2}\lambda_\alpha^{-1} \; ,
\end{eqnarray}
where the second equality follows from Eq.~(\ref{eq:Omega_eig}) \cite{Tyl18a}. 

Until now we have introduced global topological indices and local centralities
expressed through resistance distances of the original coupling network. In the upcoming sections, we show how resistance distances naturally come out when quantifying robustness of network-coupled oscillators, but that new distance metrics related to powers of the Laplacian matrix also emerge. We therefore generalize Eqs.~\eqref{eq:rdist}--\eqref{eq:Kirchhoff} to quantities corresponding to the $p^{\rm th}$ power $\mathbb{L}^p$ of the Laplacian matrix ($p\in \mathbb{N}$). This matrix is still a Laplacian matrix, and the associated 
resistance distance is defined as
\begin{eqnarray}
\Omega_{ij}^{(p)}=({\mathbb{L}^p})^\dagger_{ii} + ({\mathbb{L}^p})^\dagger_{jj} - ({\mathbb{L}^p})^\dagger_{ij} - ({\mathbb{L}^p})^\dagger_{ji} \; .
\end{eqnarray}
Still using the eigenvectors and eigenvalues of $\mathbb{L}$ we have,
\begin{eqnarray}
\Omega_{ij}^{(p)}=\sum_{\alpha \ge 2}\frac{(u_{\alpha,i}-u_{\alpha,j})^2}{\lambda_\alpha^p} \; .
\end{eqnarray} 
One can easily check that $\Omega_{ij}^{(p)}$ is still a graph-theoretic distance metric satisfying the properties mentioned between Eqs.~(\ref{eq:Omega_eig}) and (\ref{eq:C1}). We finally have generalized resistance centralities \cite{Tyl18b}
\begin{align}
\begin{split}\label{eq:Cp}
C_p(k)=\left[ n^{-1}\sum_j \Omega_{kj}^{(p)} \right]^{-1}=\left[ \sum_{\alpha\ge 2}\frac{u_{\alpha,k}^2}{\lambda_\alpha^p} + n^{-2}\Kf_p \right]^{-1}\; ,
\end{split}
\end{align}
and generalized Kirchhoff indices \cite{Tyl18a}
\begin{eqnarray}\label{eq:Kfp}
\Kf_p&=&\sum_{i<j}\Omega_{ij}^{(p)}=n\sum_{\alpha\ge 2}\lambda_\alpha^{-p} \; .
\end{eqnarray}
We note that generalized resistance distances can in principle be expressed as function of $\Omega_{ij}^{(1)}$. For instance one has
\begin{eqnarray}
\Kf_2=\frac{n}{4}\sum_{i,j}{\Omega_{ij}^{(1)}}^2-\frac{1}{2}\sum_{i,j,k}\Omega_{ij}^{(1)}\Omega_{jk}^{(1)} + \frac{(\Kf_1)^2}{n} \, .
\end{eqnarray}
Below we show how global robustness and local vulnerabilities quantified with performances measures can be expressed in terms of the resistance distance-based centralities and the generalized Kirchhoff 
indices just introduced.

\section{Synchronized oscillators under external perturbations}\label{section3}
\subsection{The Kuramoto model with inertia and its linearization}

We consider a set of network-coupled oscillators defined by 
the following set of coupled differential equations,
\begin{eqnarray}\label{eq:kuramoto}
m_i\ddot{\theta}_i+d_i\dot{\theta}_i=P_i-\sum_j b_{ij}\sin(\theta_i-\theta_j) \, .
\end{eqnarray}
Oscillators labeled $i=1,...,n$ sit on the $n$ nodes of a weighted graph 
defined by the adjacency matrix with elements $b_{ij} \ge 0$. They 
have compact angle coordinates $\theta_i\in (-\pi,\pi]$, natural frequencies $P_i/d_i$~\cite{remark1}
and inertia as well as damping parameters $m_i$ and $d_i$.  
For $m_i=0$, Eq.~\eqref{eq:kuramoto} gives the celebrated Kuramoto model on a complex network,
for which it is known that when the couplings are sufficiently strong, 
a finite fraction of, or all oscillators synchronize, i.e. with $\dot{\theta}_i-\dot{\theta}_j=0$, 
depending on the distribution of the natural 
frequencies~\cite{Kur75,Pik03,Jad04,Ace05}.
Here, we consider $P_i$ defined on a bounded, real interval and set 
$\sum_i P_i=0$ without loss of generality, 
so that synchronous states have $\dot{\theta}_i=0$, $\forall i$. 

Eq.~(\ref{eq:kuramoto}) is governed by three sets of time scales. The first one 
consists of the inverse natural frequencies $d_i/P_i$.
The second one 
is given by ratios $m_i/d_i$ and corresponds to the relaxation time of individual oscillators. 
Finally, the third one is given by the network relaxation times $d_i/\lambda_\alpha$ defined by 
the damping parameters and the eigenvalues $\lambda_\alpha$ of the weighted Laplacian matrix defined in Eq.~(\ref{eq:laplacian}) below. The first of these sets essentially determines the 
synchronous state, together with the coupling network. Depending on the other two sets of 
time scales, perturbations are locally damped or they propagate across the network~\cite{Pag19}.

We consider a stable fixed-point solution ${\bm \theta}^{(0)}=(\theta_1^{(0)},\ldots ,\theta_n^{(0)}) $ to Eq.~(\ref{eq:kuramoto}) with
unperturbed natural frequencies $\bm{P}^{(0)}$. We subject this state to a time-dependent perturbation
${\bm{P}}(t) = {\bm{P}}^{(0)} + \delta {\bm{P}}(t)$, which renders angles  time-dependent, 
${\bm{\theta}}(t) = {\bm{\theta}}^{(0)} + \delta {\bm{\theta}}(t)$. Linearizing the dynamics of Eq.~(\ref{eq:kuramoto}) about 
$\bm{\theta}^{(0)} $, one obtains  
\begin{align}\label{eq:kuramoto_lin}
{\bm M} \, \delta \ddot{\bm \theta} + {\bm D} \, \delta \dot{\bm \theta} &= \delta{\bm P}(t) - {\mathbb L}(\{ \theta_i^{(0)} \}) \, \delta {\bm \theta} \, ,
\end{align}
where we introduced inertia and damping matrices, ${\bm M}={\rm diag}\{m_i\}$ and ${\bm D}={\rm diag}\{d_i\}$, respectively, and the weighted Laplacian matrix ${\mathbb L}(\{ \theta_i^{(0)} \})$ with matrix elements
\begin{equation}\label{eq:laplacian}
{\mathbb L}_{ij} = 
\left\{ 
\begin{array}{cc}
-b_{ij} \cos(\theta_i^{(0)} - \theta_j^{(0)}) \, , & i \ne j \, , \\
\sum_k b_{ik} \cos(\theta_i^{(0)} - \theta_k^{(0)}) \, , & i=j \, .
\end{array}
\right.
\end{equation}
This Laplacian is minus the stability matrix of the linearized dynamics, and 
since we consider a stable synchronous state, it
is positive semidefinite, with a single eigenvalue $\lambda_1=0$ with eigenvector ${\bf u}_1=(1,1,1,...1)/\sqrt{n}$, and $\lambda_\alpha>0$, $\alpha=2,3,...n$. From here on, we order the 
Lyapunov exponents $\lambda_\alpha$ in increasing order, i.e. 
$\lambda_1=0< \lambda_2 < \ldots < \lambda_n$.

Eq.~(\ref{eq:kuramoto_lin}) can be solved analytically through a spectral expansion if
(i) both $\bm M$ and $\bm D$ commute with $\mathbb{L}$ or (ii) if ${{\bm M}^{-1}{\bm D}}=\gamma\mathbb{I}$. In case (i), the spectral expansion is over the
eigenmodes of $\mathbb{L}$, while in case (ii) it is over the eigenmodes of ${\bm M^{-1/2}}\mathbb{L}{\bm M^{-1/2}}$~\cite{Pag17,Col17b}. Here, we focus on case (i) with $m_i=m$, $d_i=d$ $\forall i$. 

Expanding the angle deviations over the eigenmodes of $\mathbb{L}$ as $\delta {\bm \theta}(t)=\sum_{\alpha}c_\alpha(t) {\bf u}_\alpha$, Eq.~(\ref{eq:kuramoto_lin}) leads to a Langevin equation,
\begin{eqnarray}
m\; \ddot{c}_\alpha(t) + d\; \dot{c}_\alpha(t) = \delta {\bm P}(t)\cdot {\bf u}_\alpha -\lambda_\alpha \; c_\alpha(t) \; ,
\end{eqnarray}
whose general solution reads
\begin{equation}\label{eq:calpha}
\begin{aligned}
c_{\alpha}(t)=&m^{-1} \, e^{-(\gamma+\Gamma_{\alpha})t/2}\int_0^{t}e^{{\Gamma_{\alpha}}t_1}\\
&\times \int_{0}^{t_1}\delta {\bm{P}}(t_2)\cdot {\bf{u}}_{\alpha} \, e^{(\gamma-\Gamma_{\alpha}) t_2 /2} \, {\rm d}t_2{\rm d}t_1 \;,
\end{aligned}
\end{equation}
with $\Gamma_\alpha=\sqrt{\gamma^2-4\lambda_\alpha/m}$ and $\gamma=d/m$. Similar expressions have been derived using the transfer function formalism \cite{Pag17,Guo18} or within linear response \cite{Man17b,Tyl18a,Ket16}.
When $\gamma^2<4\lambda_\alpha/m$, $\Gamma_\alpha \in i \mathbb{R}$ and accordingly, 
$|\Gamma_\alpha|$ corresponds to the angular frequency of oscillations along 
the eigenmode ${\bf u}_\alpha$ of $\mathbb{L}$. When on the other hand 
$\gamma^2>4\lambda_\alpha/m$, $\Gamma_\alpha \in \mathbb{R}$ and gives an additional 
damping beyond $\gamma$. From Eq.~\eqref{eq:calpha}, angle and frequency deviations 
can be calculated as $\delta {\bm \theta}(t)=\sum_{\alpha}c_\alpha(t){\bf u}_\alpha$.

\subsection{Performance Measures}

The perturbation $\delta {\bm P}(t)$ moves the 
oscillators angles and frequencies away from their value at synchrony and renders them time
dependent. For not too strong, finite-time perturbations, they eventually relax to their synchronous values
and to assess the magnitude of the 
excursion away from synchrony, we introduce the following quadratic performance measures 
\begin{subequations}\label{eq:c12}
\begin{eqnarray}\label{eq:c1}
{\mathcal P}_1(T) &=& \sum_i \int_0^T \, |\delta \theta_i(t) - \Delta(t) |^2 {\rm d}t  \, , \\ \label{eq:c2}
{\mathcal P}_2(T) &=& \sum_i  \int_0^T \,  |\delta \dot{\theta}_i(t)- \dot \Delta(t) |^2 {\rm d}t   \, .
\end{eqnarray}
\end{subequations}
Similar measures have been discussed in the context of consensus algorithms \cite{Bam12,Gru18}, electric power systems \cite{Poo17,Pag17,Sia16} and coupled oscillators systems \cite{Tyl18a,Tyl18b}.
The results we are about to present directly connect 
these performance measures to resistance-distance
based centralities and Kirchhoff indices introduced in Section~\ref{section2}. While similar connections
may have been inferred from some of these works 
(in particular Refs.~\cite{Pag17,Sia16,Tyl18a}),
to the best of our knowledge, it was first unambiguously stated in Ref.~\cite{Tyl18b}. 

Because synchronous states are defined modulo any
homogeneous angle shift, they are unaffected by the transformation $\theta_i^{(0)} \rightarrow \theta_i^{(0)} + C$. Accordingly, only angle shifts with $\sum_i \delta \theta_i(t)=0$ matter, which is incorporated
in the definitions of ${\mathcal P}_{1,2}$ by subtracting averages $\Delta (t) = n^{-1}
\sum_j \delta \theta_j(t)$ and $\dot\Delta (t)  = n^{-1} \sum_j \delta \dot\theta_j(t)$. 
If the perturbation is not too strong and finite in time, both $\mathcal{P}_1$ and $\mathcal{P}_2$ are finite even for $T \rightarrow \infty$. 
Low values for ${\mathcal P}_{1,2}^\infty \equiv {\mathcal P}_{1,2}(T\rightarrow \infty)$ indicate then that the system absorbs the perturbation with little fluctuations, while large values
indicate a temporary fragmentation of the system into independent pieces -- qualitatively speaking,
${\mathcal P}_{1,2}^\infty$ measures
the coherence of the synchronous state~\cite{Bam12}. 

Using the spectral expansion with coefficients given in Eq.~\eqref{eq:calpha}, the
performance measures of Eqs.~(\ref{eq:c12})
read, in our case of homogeneous inertia and damping coefficients
\begin{subequations}
\begin{eqnarray}
{\mathcal P}_1(T) &=& \sum_{\alpha\ge 2} \int_0^T \, c^2_\alpha(t) {\rm d}t \; , \\
{\mathcal P}_2(T) &=& \sum_{\alpha\ge 2} \int_0^T \, \dot{c}^2_\alpha(t) {\rm d}t \; .
\end{eqnarray}
\end{subequations}
Performance measures depend on the perturbation vector $\delta \bm{P}(t) = \delta \bm{P}_0 \, f(t) $,
which may have different time dependences $f(t)$ -- such as, for instance, noisy fluctuations or instantaneous, Dirac-delta
perturbations -- and different geographical dependences encoded in $ \delta \bm{P}_0$.
In this manuscript we consider quenches where $f(t)$ vanishes outside
some time interval, inside which it is constant but nonzero. 
In the next section we calculate performance measures for general
perturbation vectors $\delta  \bm{P}_0$ for such quenches.
As for geographical dependences,
we then consider two cases of (i) nodal vulnerabilities, with
local perturbations $\delta \bm{P}_0 = (0,... ,\delta P_{0,k}, ... , 0)$ and (ii) global robustness, 
where performance measures are
averaged over all possible locations $k$ for the perturbation. 

\subsection{Quench Perturbation}

We compute both performance measures ${\mathcal P}_{1,2}$ for a quench perturbation $\delta \bm{P}(t)= \delta \bm{P}_0 \,\Theta(t) \, \Theta(\tau_0-t)$ with the Heaviside function $\Theta(t)$
and a perturbation vector $\delta {\bm P}_0$ encoding the geographical distribution of the perturbation. The duration $\tau_0$
of the quench allows to explore the different time scales of the system and we show below that
${\mathcal P}_{1,2}$ varies significantly depending on $\tau_0$.
Using Eq.~(\ref{eq:calpha}), Eqs.(\ref{eq:c12}) give
\begin{widetext}
\begin{subequations}\label{eq:c12heaviside}
\begin{eqnarray}
\mathcal{P}^\infty_1 &=&\frac{m}{8\gamma} \sum_{\alpha\ge 2}\frac{{(\delta \bm{P}_0 \cdot {\bf u}_\alpha)^2}}{\Gamma_{\alpha}\lambda_{\alpha}^3}\left[ 2\Gamma_{\alpha} ( 4\gamma\tau_0\lambda_\alpha /m - 3\gamma^2 - \Gamma_\alpha^2 ) + ( \gamma+\Gamma_\alpha )^3e^{-\tau_0 \frac{(\gamma-\Gamma_\alpha)}{2}} - ( \gamma-\Gamma_\alpha )^3e^{-\tau_0\frac{(\gamma+\Gamma_\alpha)}{2}} \right] \, , \\
\mathcal{P}^\infty_2&=& \frac{1}{2 d}\sum_{\alpha\ge 2}\frac{(\delta \bm{P}_0 \cdot {\bf u}_\alpha)^2}{\Gamma_\alpha\lambda_\alpha}\left[2\Gamma_\alpha -(\gamma+\Gamma_\alpha)e^{-\frac{\tau_0(\gamma-\Gamma_\alpha)}{2}} +(\gamma-\Gamma_\alpha)e^{-\frac{\tau_0(\gamma+\Gamma_\alpha)}{2}}  \right] \; .
\end{eqnarray}
\end{subequations} 
\end{widetext}
It is easily checked that ${\mathcal P}_{1,2}^\infty \in \mathbb{R}$ in both cases 
$\gamma^2>4\lambda_\alpha/m$ (with $\Gamma_\alpha \in \mathbb{R}$) and 
$\gamma^2<4\lambda_\alpha/m$ (with $\Gamma_\alpha \in i \mathbb{R}$).

Both performance measures are given by a spectral sum of terms corresponding to the eigenmodes
of the network Laplacian matrix $\mathbb{L}$. Each term in this sum
depends on the scalar product of the perturbation vector $\delta \bm{P}_0$ 
with the eigenmodes ${\bf u}_\alpha$ of $\mathbb L$ times a mode-dependent factor. 
The latter is an almost always decreasing function of the eigenvalues $\lambda_\alpha$. 
Therefore,  Eqs.~\eqref{eq:c12heaviside} suggest that the largest excursion can be obtained by overlapping $\delta {\bm P}_0$ with few of the lowest-lying eigenmodes of $\mathbb{L}$, in particular
${\bf u}_2$,  the so-called Fiedler mode of the network \cite{Fie73}. 

To get more insight into Eqs.~\eqref{eq:c12heaviside}, we compute their two asymptotic limits of long and short 
$\tau_0$. For perturbations with 
very short duration i.e. $\tau_0\ll m/d$, $(\gamma\pm \Gamma_\alpha)^{-1}$, we have,
\begin{subequations}\label{eq:P_s_I}
\begin{align}
\begin{split}\label{eq:P1_s_I}
\mathcal{P}^\infty_1 &=\frac{\tau_0^2}{2 d}\sum_{\alpha\ge 2}\frac{(\delta{\bm P}_0\cdot{\bf u}_{\alpha})^2}{\lambda_{\alpha}} ,
\end{split}\\
\begin{split}\label{eq:P2_s_I}
\mathcal{P}^\infty_2 &= \frac{\tau_0^2}{2m d}\sum_{\alpha\ge 2}{(\delta{\bm P}_0\cdot{\bf u}_{\alpha})^2} \; .
\end{split} 
\end{align}
\end{subequations}
Each term in the sum over modes depends 
on $\lambda_\alpha$ for $\mathcal{P}_1^\infty$ but not for $\mathcal{P}_2^\infty$.
Consequently, $\mathcal{P}_1^\infty$ depends explicitly on the location of the 
perturbation, while there is no such dependence for $\mathcal{P}_2^\infty$, which depends
only on the squared norm of the perturbation vector $\delta {\bm P}_0$ orthogonal to ${\bf u}_1$.
This reflects the fact that in the regime of short $\tau_0$, the perturbation does not  
act long enough to change the kinetic energy of inertiafull oscillators, which $\mathcal{P}_2^{\infty}$
essentially measures. Consequently, 
the perturbation is quickly damped locally, with little dependence on its location in the 
situation we consider of 
homogeneously distributed inertia.
We note that similar topology-independent results were obtained for
instantaneous, Dirac-delta perturbations \cite{Bam13}. 

In the other limit $\tau_0\gg m/d$, $(\gamma\pm \Gamma_\alpha)^{-1}$, the performance measures read
\begin{subequations}\label{eq:P_l_I}
\begin{align}
\begin{split}\label{P1_l_I}
\mathcal{P}^\infty_1 &={\tau_0}\sum_{\alpha\ge 2}\frac{(\delta{\bm P}_0\cdot{\bf u}_{\alpha})^2}{\lambda_{\alpha}^2} \, ,
\end{split}\\
\begin{split}\label{P2_l_I}
\mathcal{P}^\infty_2 &=d^{-1} \, \sum_{\alpha\ge 2}\frac{(\delta{\bm P}_0\cdot{\bf u}_{\alpha})^2}{\lambda_{\alpha}}
\end{split} \; .
\end{align}
\end{subequations}
In this case of a long-lasting perturbation, both $\mathcal{P}_1^{\infty}$ and $\mathcal{P}_2^{\infty}$ 
depend on the location of the perturbation. Furthermore, and perhaps more importantly, 
the inertia affects neither $\mathcal{P}_1^\infty$ nor $\mathcal{P}_2^{\infty}$. This is so since, 
for long quenches, oscillators have the time to synchronize at a new frequency with zero angular acceleration before the perturbation is over. 
We further note that $\tau_0$ no longer appears in $\mathcal{P}_2^{\infty}$, since the latter considers deviations orthogonal to  ${\bf u}_1$. Consequently, the whole time spent by the oscillators at the new frequency does not contribute to $\mathcal{P}_2^{\infty}$.
Most importantly, Eqs.~\eqref{eq:P_s_I} and \eqref{eq:P_l_I} suggest that in both asymptotic 
limits of short and long perturbations, $\mathcal{P}_{1,2}^{\infty} \propto \sum_{\alpha\ge 2}(\delta{\bm P}_0\cdot{\bf u}_{\alpha})^2 / \lambda_{\alpha}^p$ with $p=0,1,2$. That result was already hinted at in Ref.~\cite{Tyl18a} for inertialess oscillators and various types of perturbations. Below we show how
this dependence leads to performance measures depending on the resistance distances, 
centralities and Kirchhoff indices introduced in Section~\ref{section2}.

Eqs.~\eqref{eq:c12heaviside} and their asymptotic limits, Eqs.~\eqref{eq:P_s_I} and \eqref{eq:P_l_I},
give the performance measures  $\mathcal{P}_{1,2}^\infty$ for any perturbation 
vector $\delta {\bm P}_0$. We next discuss two important cases of 
(i) a single-node perturbation, $\delta P_{0,i}=\delta P_0 \delta_{ik}$, where large values of
the node-dependent performance measures
$\mathcal{P}_{1,2}^\infty \rightarrow \mathcal{P}_{1,2}^\infty(k)$ identify local
vulnerabilities and (ii) averaged perturbation over ensemble of homogeneously distributed 
perturbation vectors $\delta {\bm P}_0$, where large values of
$\mathcal{P}_{1,2}^\infty \rightarrow \langle \mathcal{P}_{1,2}^\infty \rangle$ indicate
globally fragile networks. 

\subsection{Specific Local Vulnerabilities}
To assess local vulnerabilities of the coupled oscillators, we apply a quench perturbation on a single node. The vulnerability of node $k$ is then given by Eqs.~(\ref{eq:c12heaviside}) with  
 the components of the perturbation vector given by 
 $\delta P_{0,i}=\delta P_0 \delta_{ik}$. In the limit of short duration of perturbation,  $\tau_0\ll m/d$, $(\gamma\pm \Gamma_\alpha)^{-1}$, one obtains
\begin{subequations}\label{P_s}
\begin{align}
\begin{split}\label{eq:P1_s}
\mathcal{P}^\infty_1(k) &=\frac{\delta P_0^2\tau_0^2}{2d}\sum_{\alpha\ge 2}\frac{{{u}_{\alpha,k}^2}}{\lambda_{\alpha}}=\frac{\delta P_0^2\tau_0^2}{2d}[C_1^{-1}(k)-n^{-2}\Kf_1] \, ,
\end{split}\\\label{eq:P2_s}
\mathcal{P}^\infty_2(k) &= \frac{\delta{P}_0^2\tau_0^2}{2md}\sum_{\alpha\ge 2}{{u}_{\alpha,k}^2}=\frac{\delta{P}_0^2\tau_0^2}{2md}\frac{(n-1)}{n} \; ,
\end{align}
\end{subequations}
where the right-hand side of Eq.~\eqref{eq:P1_s} directly follows from Eq.~\eqref{eq:C1}.
For a perturbation on node $k$, $\mathcal{P}_1^\infty(k)$ is expressed in terms of the centrality,
$C_1(k)$, a local nodal descriptor, and the Kirchhoff index $\Kf_1$, a 
global network descriptor. Consequently, the most vulnerable nodes in a given network, 
according to
$\mathcal{P}_1^\infty(k)$, are identified by their resistance-distance based centrality $C_1(k)$.

In the other limit of long perturbations, $\tau_0\gg m/d$, $(\gamma\pm \Gamma_\alpha)^{-1}$, Eqs.~(\ref{eq:P_l_I}) give
\begin{subequations}\label{P_l}
\begin{align}
\begin{split}\label{P1_l}
\mathcal{P}^\infty_1(k) &={\delta P_0^2 \tau_0}\sum_{\alpha\ge 2}\frac{{{u}_{\alpha,k}^2}}{\lambda_{\alpha}^2}=\delta P_0^2\tau_0[C_2^{-1}(k)-n^{-2}\Kf_2] \, ,
\end{split}\\
\begin{split}\label{P2_l}
\mathcal{P}^\infty_2(k) &=\frac{\delta P_0^2}{d}\sum_{\alpha\ge 2}\frac{{{u}_{\alpha,k}^2}}{\lambda_{\alpha}}=\frac{\delta P_0^2}{d}[C_1^{-1}(k)-n^{-2}\Kf_1] 
\end{split} \; .
\end{align}
\end{subequations}
This time $\mathcal{P}_1^{\infty}$ is given by the higher order centrality $C_2(k)$ and 
Kirchhoff index $\Kf_2$.

When considering a given, fixed network,
Eqs.~\eqref{P_s} and \eqref{P_l} show that perturbations on 
the most central nodes -- as measured by either centrality $C_1$ or $C_2$ -- 
give the smallest overall responses,
except when considering $\mathcal{P}^\infty_2(k)$ for a short-time perturbation. In that latter case,
the response is homogeneous and perturbing any node leads to the same performance 
measure $\mathcal{P}^\infty_2(k)$. 
When comparing two nodes with similar centrality on two different networks, on the other hand, 
Eqs.~\eqref{P_s} and \eqref{P_l} indicate that the largest response occurs on the network with 
smallest generalized Kirchhoff index -- except again for $\mathcal{P}^\infty_2(k)$ and a short-time perturbation. We show below that the overall 
network robustness is actually given by these generalized Kirchhoff indices, which makes
this observation quite counterintuitive : when perturbing two nodes of equal centrality on two 
different networks, the
largest response is actually recorded on the overall more robust network !
We will come back to this point below.

\subsection{Averaged Global Robustness}
We next assess the global robustness of synchrony in a given network, by 
averaging Eqs.~(\ref{eq:c12heaviside}) over an homogeneously distributed 
ensemble of perturbation vectors defined by $\langle\delta P_{0,i}\delta P_{0,j}\rangle=\delta_{ij}\langle \delta P_0^2\rangle$ \cite{Tyl18a}. Averaging Eqs.~(\ref{eq:c12heaviside}) gives, in the limit
of short perturbations,  $\tau_0\ll m/d$, $(\gamma\pm \Gamma_\alpha)^{-1}$
\begin{subequations}\label{eq:gl_p_s}
\begin{eqnarray}\label{eq:gl_p1_s}
\langle \mathcal{P}_1^{\infty} \rangle &=& \frac{\langle \delta P_0^2 \rangle\tau_0^2}{2d}\sum_{\alpha\ge 2}\lambda_\alpha^{-1}=\frac{\langle \delta P_0^2 \rangle\tau_0^2}{2nd}\Kf_1 \; ,\\
\langle \mathcal{P}_2^{\infty} \rangle &=& \frac{\langle \delta P_0^2 \rangle\tau_0^2}{2md}\frac{n-1}{n} \; .\label{eq:gl_p2_s}
\end{eqnarray}
\end{subequations}
We see that $\langle \mathcal{P}_1^{\infty} \rangle$ is given by the Kirchhoff index $\Kf_1$ which is proportional to the network's average resistance distance $\Omega_{ij}^{(1)}$ 
[see Eq.~\eqref{eq:Kirchhoff}]. Similarly to the local vulnerability in this limit, $\langle \mathcal{P}_2^{\infty} \rangle$ depends on the network only marginally through the number of nodes.

\begin{figure}[h!]
\includegraphics[width=0.91\columnwidth]{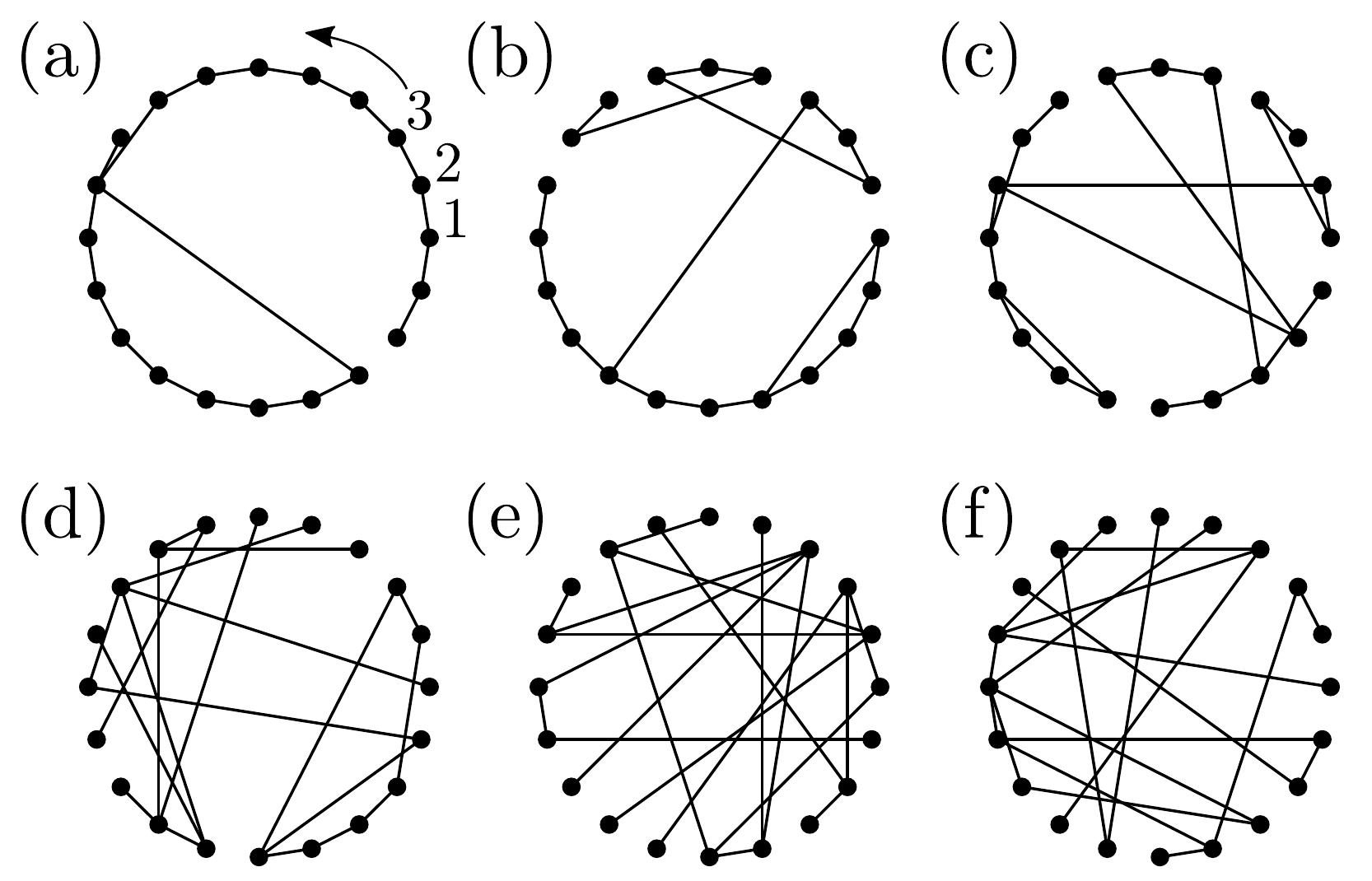}
\caption{Six networks with $n=20$ nodes obtained by the rewiring procedure of Ref.~\cite{Wat98},
starting from a cyclic graph and rewiring every edge of the network with a probability $p=0.15$ (a), $p=0.3$ (b), $p=0.45$ (c), $p=0.6$ (d), $p=0.75$ (e) and $p=0.9$ (f). The node numbering
used in Fig.~\ref{fig2} is indicated in panel (a).}\label{fig1}
\end{figure}

In the other limit $\tau_0\gg m/d$, $(\gamma\pm \Gamma_\alpha)^{-1}$, Eqs.~(\ref{eq:P_l_I}) give
\begin{subequations}\label{eq:gl_p_l}
\begin{eqnarray}
\langle \mathcal{P}_1^{\infty} \rangle &=& {\langle \delta P_0^2 \rangle \tau_0}\sum_{\alpha\ge 2}\lambda_\alpha^{-2}=\frac{\langle \delta P_0^2 \rangle \tau_0}{n}\Kf_2 \; ,\label{eq:gl_p1_l}\\
\langle \mathcal{P}_2^{\infty} \rangle &=& \frac{\langle \delta P_0^2 \rangle}{d}\sum_{\alpha\ge 2}\lambda_\alpha^{-1}=\frac{\langle \delta P_0^2 \rangle}{nd}\Kf_1 \; .\label{eq:gl_p2_l}
\end{eqnarray}
\end{subequations}
Both performance measures depend on generalized Kirchhoff indices. Quite remarkably and as for local vulnerabilities, the only average performance measure that depends
on inertia is $\langle \mathcal{P}_2^{\infty}\rangle$ in the short $\tau_0$ limit. 
In the next Section, we numerically confirm the validity of the 
analytical theory presented in this Section.

\section{Numerical Results}\label{section4}
\subsection{Local Vulnerabilities and Resistance Centralities}

We numerically investigate local vulnerabilities by perturbing individual nodes with the quench perturbation discussed above.
Our theory applies to network of any geometry with any number $n$ of nodes. However in order to 
better visualize the agreement between analytical predictions and numerical results we restrict ourselves to relatively small graphs with $n=20$ nodes of the kind shown in Fig.~\ref{fig1}. 
\begin{figure}[H]
\centering
\includegraphics[scale=0.35]{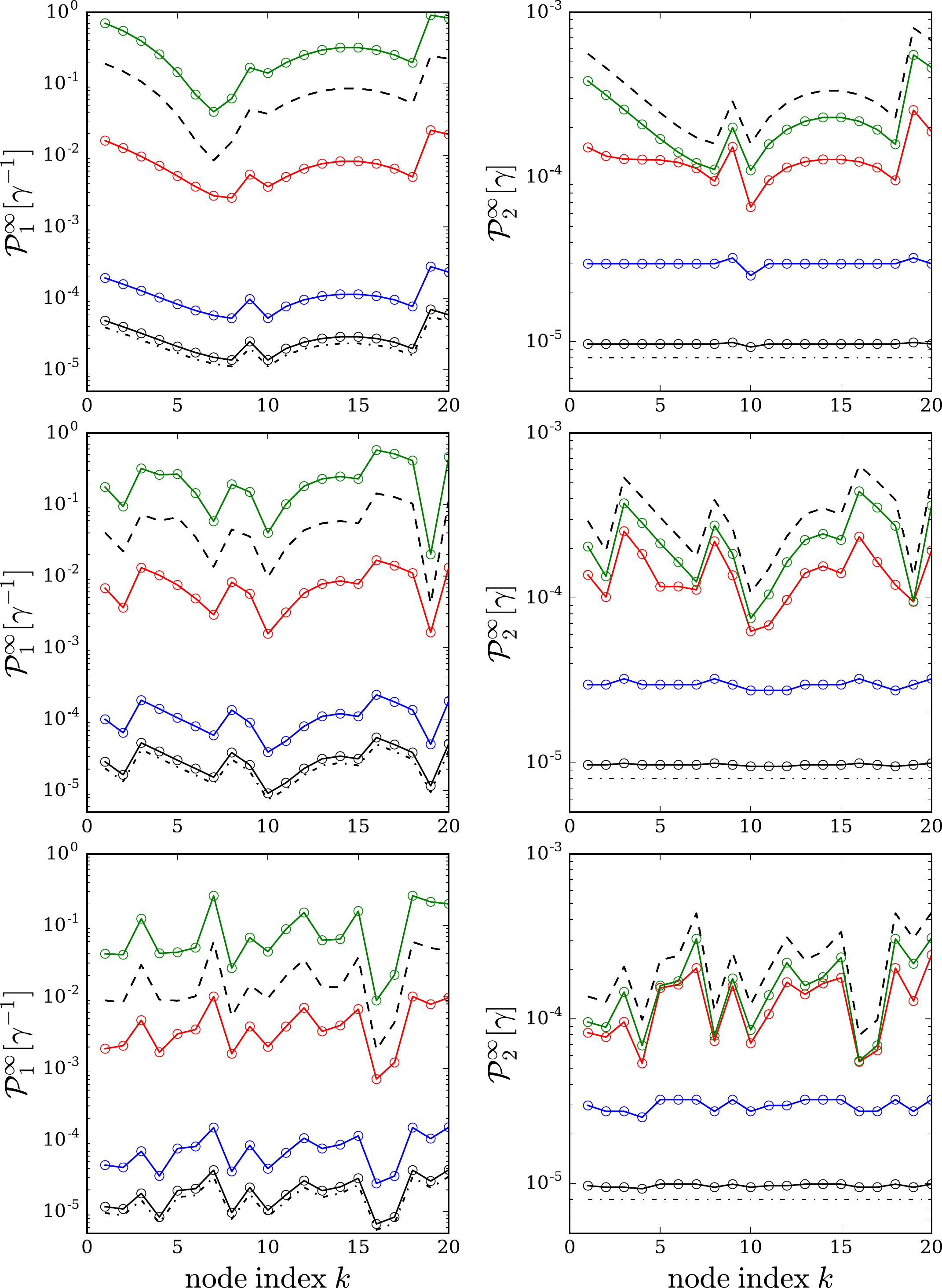}
\caption{Performances measures $\mathcal{P}_1$ (left) and $\mathcal{P}_2$ (right) for the graphs of Fig.~\ref{fig1}a (top), Fig.~\ref{fig1}c, (middle), Fig.~\ref{fig1}e (bottom) and a quench perturbation of magnitude $\delta P_0=0.01$ on node $k$. Numerical results (circles) and analytical Eqs.~(\ref{eq:c12heaviside}) (solid lines) are plotted for different durations of perturbation $\gamma \tau_0=0.5$ (black), $1$ (blue), $10$ (red), $100$ (green). The asymptotic values of short and long 
$\tau_0$ given in 
Eqs.~(\ref{P_s}) (dotted line) and (\ref{P_l}) (dashed line) are shown, vertically shifted by an arbitrary amount  for clarity. 
The node numbering is given in Fig.~\ref{fig1}a.}\label{fig2}
\end{figure}
We check Eqs.~(\ref{eq:c12heaviside}) for the model defined in Eq.~(\ref{eq:kuramoto}) 
with $b_{ij}=1$ on the edge of the graph considered and 
$b_{ij}=0$ otherwise, $m_i\equiv m=1$ and $d_i\equiv d=1$. We numerically time-evolve Eq.~(\ref{eq:kuramoto})  with a fourth-order 
Runge-Kutta method,
following a perturbation $\delta {P_{i}}(t)= \delta {P}_0\delta_{ik} \,\Theta(t) \, \Theta(\tau_0-t)$  
away from ${\bm P}^{(0)}=0$ and
starting from the corresponding synchronous state ${\bm \theta}^{(0)}=0$. Fig.~\ref{fig2} shows that the theory of Eqs.~(\ref{eq:c12heaviside}) is in perfect agreement with numerical results. In particular, one clearly sees the crossover from $[C_1^{-1}(k)-n^{-2}\Kf_1]$ to $[C_2^{-1}(k)-n^{-2}\Kf_2]$ for $\mathcal{P}_1^\infty$ 
(dotted to dashed lines on the left panels) and from a constant to $[C_1^{-1}(k)-n^{-2}\Kf_1]$ (dotted to dashed line on the right panels) for $\mathcal{P}_2^\infty$, as $\tau_0$ increases.
This  fully confirms our theoretical predictions, Eqs.~(\ref{P_s})-(\ref{P_l}).
We conclude that,
generally speaking (i.e. except for $\mathcal{P}_2^{\infty}$ and short perturbations), the most central nodes are the most robust. They are connected by multiple paths to the rest of the network, and when they are perturbed, the disturbance quickly diffuses across the network with small angle differences. In contrast, the most peripheral nodes such as dead ends have only few paths connecting them to the bulk of the network and the disturbance diffuses across the network with large angle differences. 
It has been numerically found that dead ends  undermine grid stability~\cite{Men14}, and our results shed some analytical light on that observation. 

\begin{figure}
\includegraphics[width=0.93\columnwidth]{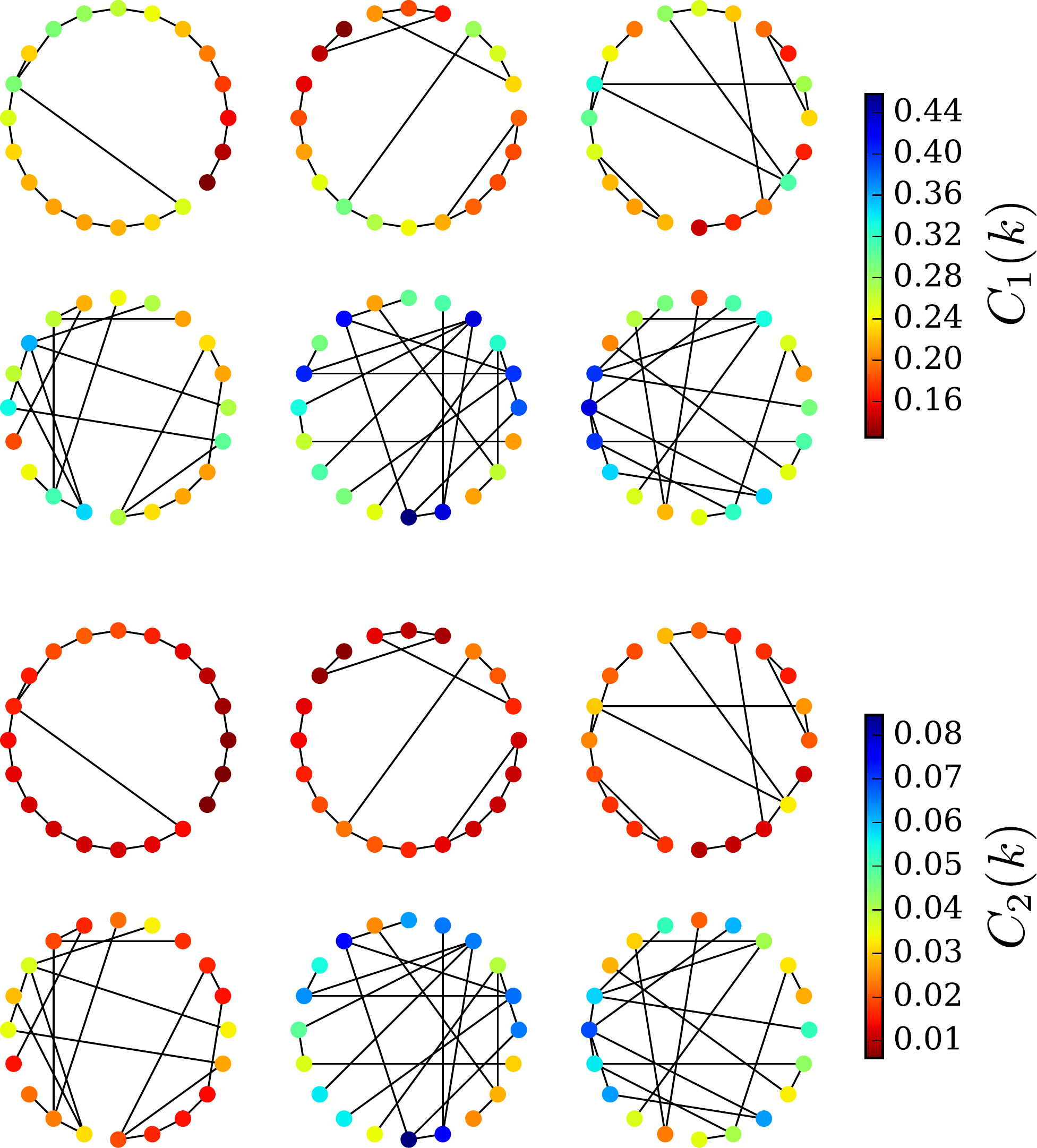}
\caption{Resistance centralities $C_1(k)$ (top) and $C_2(k)$ (bottom), given in Eqs.~(\ref{eq:C1}) and (\ref{eq:Cp}) respectively, for the six graphs of Fig.~\ref{fig1}.}\label{fig3}
\end{figure}

\begin{figure}[H]
\includegraphics[width=0.98\columnwidth]{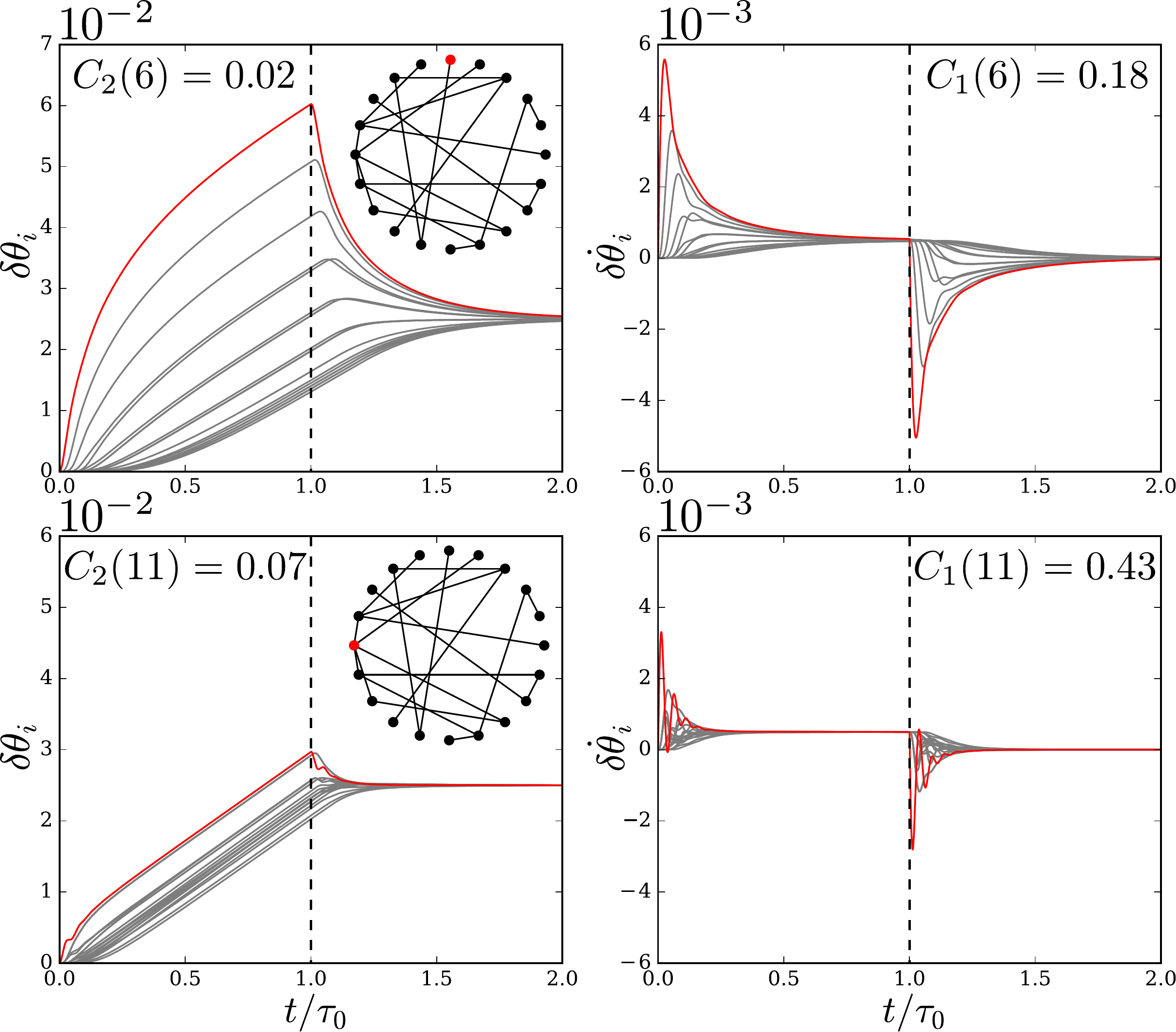}
\caption{Time-evolution of angles (left) and frequencies (right) following a quench perturbation applied on node $6$ (top panels) and $11$ (bottom panels) of graph (f) in Fig.~\ref{fig1} with $\gamma\tau_0=50$. The trajectory of the perturbed oscillator is shown in red. Angles and frequencies spread more when the perturbation is applied on node $6$ than on node $11$,  in agreement with predictions of Eqs.~(\ref{P_l}) since node $6$ has the smallest, node $11$ the largest centrality in this
graph.}\label{fig32}
\end{figure}

We further illustrate this strong connection between resistance centralities and response of the system. We show in Fig.~\ref{fig3} resistance centralities $C_1(k)$ and $C_2(k)$ for the six graphs of Fig.~\ref{fig1}. One sees that $C_1(k)$ and $C_2(k)$ tend to become higher while going from graph (a) to (f) indicating that graphs with more rewired edges (and thus with more long-range couplings) have shorter distances between nodes and thus lower Kirchhoff indices. Interestingly,
several nodes with a high centrality $C_1(k)$ do not necessarily have a high centrality $C_2(k)$,
and vice-versa. We then show in Fig.~\ref{fig32} the time-evolution of angles and frequencies 
following a local quench perturbation on two different nodes of graph (f) with very different resistance centralities. One clearly sees that for a perturbed node with low resistance centrality (Fig.~\ref{fig32}, top), angles and frequencies spread more during the perturbation than for a node with higher centrality (Fig.~\ref{fig32}, bottom). 

\begin{figure*}
\includegraphics[width=0.98\textwidth]{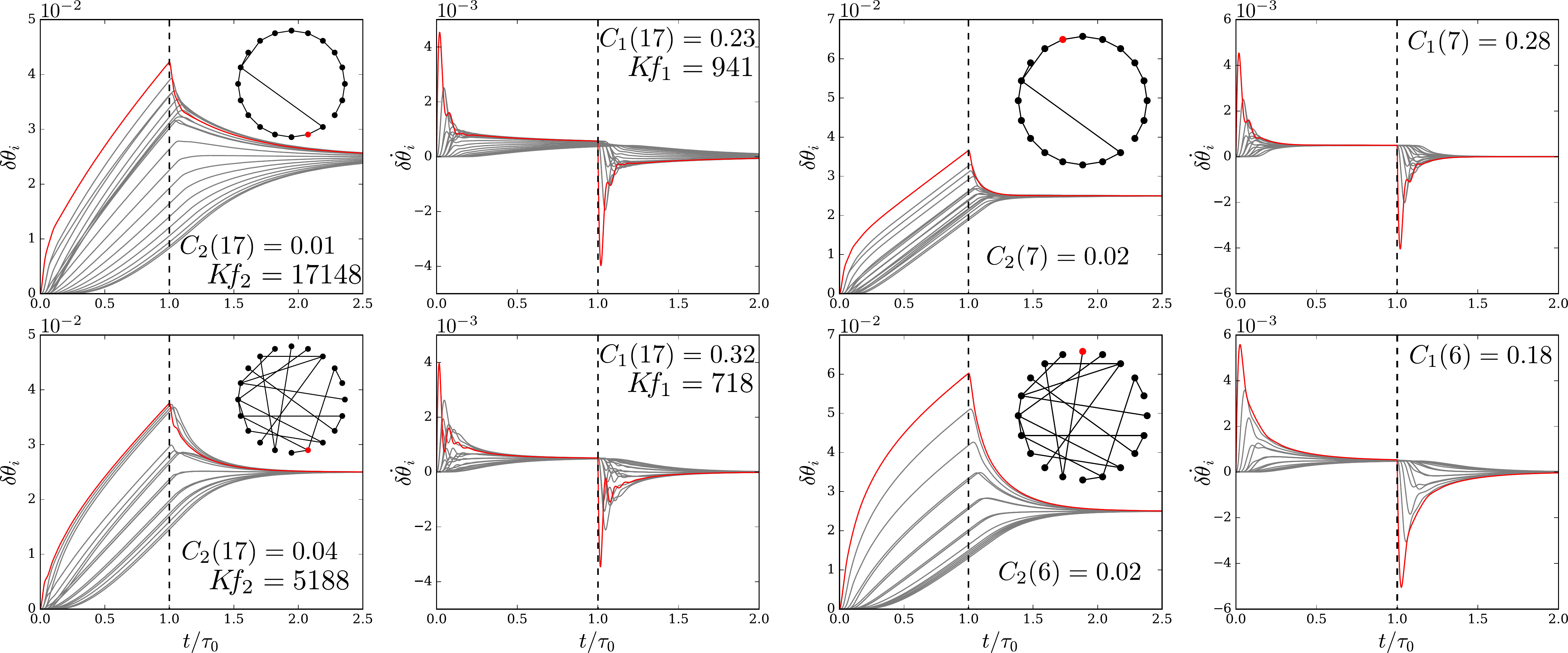}
\caption{Trajectories of angles and phases for the graphs of Fig.~\ref{fig1} with $p=0.15$ (top) and $p=0.9$ (bottom) obtained by numerically time-evolving Eq.~(\ref{eq:kuramoto}) for the same quench perturbation with $\gamma\tau_0=50$ applied on the node colored in red in the insets. In the four left panels, perturbed nodes are close to median value of $C_1(k)$, respectively in graph with $p=0.15$ (top) and $p=0.9$ (bottom). In the four right panels, perturbed nodes are the most (top) and least (bottom) central ones according to $C_2(k)$, respectively in graph with $p=0.15$ and $p=0.9$.}\label{fig5}
\end{figure*}

Generally speaking, networks with higher rewiring probabilities have smaller global topological indices $\Kf_1$ and $\Kf_2$ and thus smaller $\langle\mathcal{P}_{1,2}^{\infty}\rangle$ according to our theory. This is confirmed numerically in the four left panels in Fig.~\ref{fig5}, where we apply the same quench perturbation on nodes with resistance centralities $C_1(k)$ close to their median value in the
corresponding graph. One observes that angles and frequencies spread more and take more time to return to the initial fixed point in the graph with higher $\Kf_1$ and $\Kf_2$ (top) compared to the one with more rewired edges (bottom). 

While this is a rather general rule, it does not forbid
exceptions. As a matter of fact, specific perturbations can lead to 
higher response in a network with lower Kirchhoff index than in a network with higher Kirchhoff index. 
Such an exception is illustrated in the four right panels in Fig.~\ref{fig5}, where the same quench perturbation is applied on nodes with similar resistance centralities $C_1(k)$ but belonging to graphs with very different Kirchhoff indices (see insets of Fig.~\ref{fig5}). 
As expected from Eqs.~(\ref{P_l}), if two nodes on different networks 
have the same centralities, then, a perturbation applied
on the one in the network with lower Kirchhoff index produces the largest response. Another illustration of this effect is given in Fig.~\ref{fig1_intro}, where graph (e) is more robust than graph (a) on average (dashed lines). But if we compare the response to specific local perturbations, some nodes of graph (a) are more robust than those of graph (e) (red crosses). Both the generic and the exceptional behaviors are accurately captured by our theory.

\subsection{Global Robustness and Generalized Kirchhoff Indices}
\begin{figure}
\includegraphics[scale=0.4]{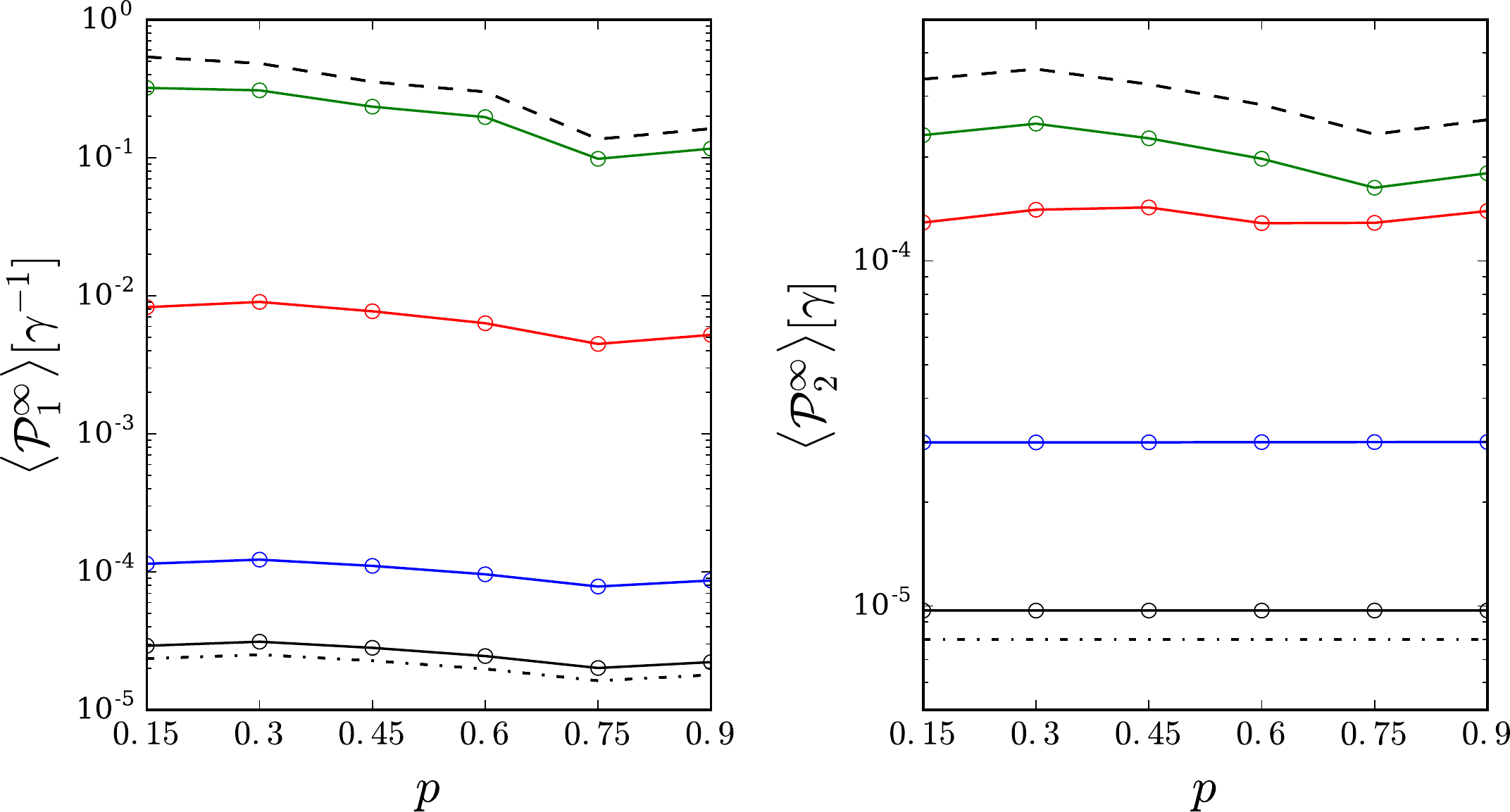}
\caption{Averaged Performances measures $\langle\mathcal{P}_1^\infty\rangle$, $\langle\mathcal{P}_2^\infty\rangle$ for the graphs of Fig.~\ref{fig1} obtained numerically (circles) and predicted 
analytically, Eqs.~(\ref{eq:c12heaviside}) (solid lines) for perturbations with $\gamma\tau_0=0.5$ (black), $1$ (blue), $10$ (red), $100$ (green). The asymptotic values of short and long 
$\tau_0$ given in 
Eqs.~(\ref{eq:gl_p_s}) (dotted line) and (\ref{eq:gl_p_l}) (dashed line) are shown, vertically shifted by an arbitrary amount  for clarity.}\label{fig4}
\end{figure}

We next investigate global robustness by averaging performance measures over
an ensemble of perturbation vectors located on a single node, $\delta {\bm P}_0=(0,..,\delta P_{0,k},0,...)$ with $k=1,...,n$. 
Fig.~\ref{fig4} compares the resulting numerical averages $\langle \mathcal{P}_{1,2}^\infty\rangle$ with the average of the theoretical prediction of Eqs.~(\ref{eq:c12heaviside}). 
Numerics and theory agree well. In particular the left panel confirms nicely the crossover between $\Kf_1$ and $\Kf_2$ predicted by Eqs.~(\ref{eq:gl_p1_s}) and (\ref{eq:gl_p1_l}). 
A similar behavior is visible in the right panel, where 
$\langle \mathcal{P}_{2}^\infty\rangle$ does not depend on the network topology 
for short duration of perturbation (black and blue lines and symbols) but 
crosses over to $\Kf_1$ as $\tau_0$ increases, as predicted by Eqs.~(\ref{eq:gl_p2_s}) and (\ref{eq:gl_p2_l}).

We finally note that networks with high $\Kf_1$ do not necessarily have a high $\Kf_2$,
and vice-versa.  This is illustrated in Fig.~\ref{fig4} where the chosen network with $p=0.15$ has a higher $\Kf_2$ but a lower $\Kf_1$ than the chosen network with $p=0.3$.
Below we analyze in more details $\Kf_{1,2}$ in randomly rewired networks.

\subsection{Generalized Kirchhoff Indices in Small-World Networks}

\begin{figure}[H]
\begin{center}
\includegraphics[width=0.95\columnwidth]{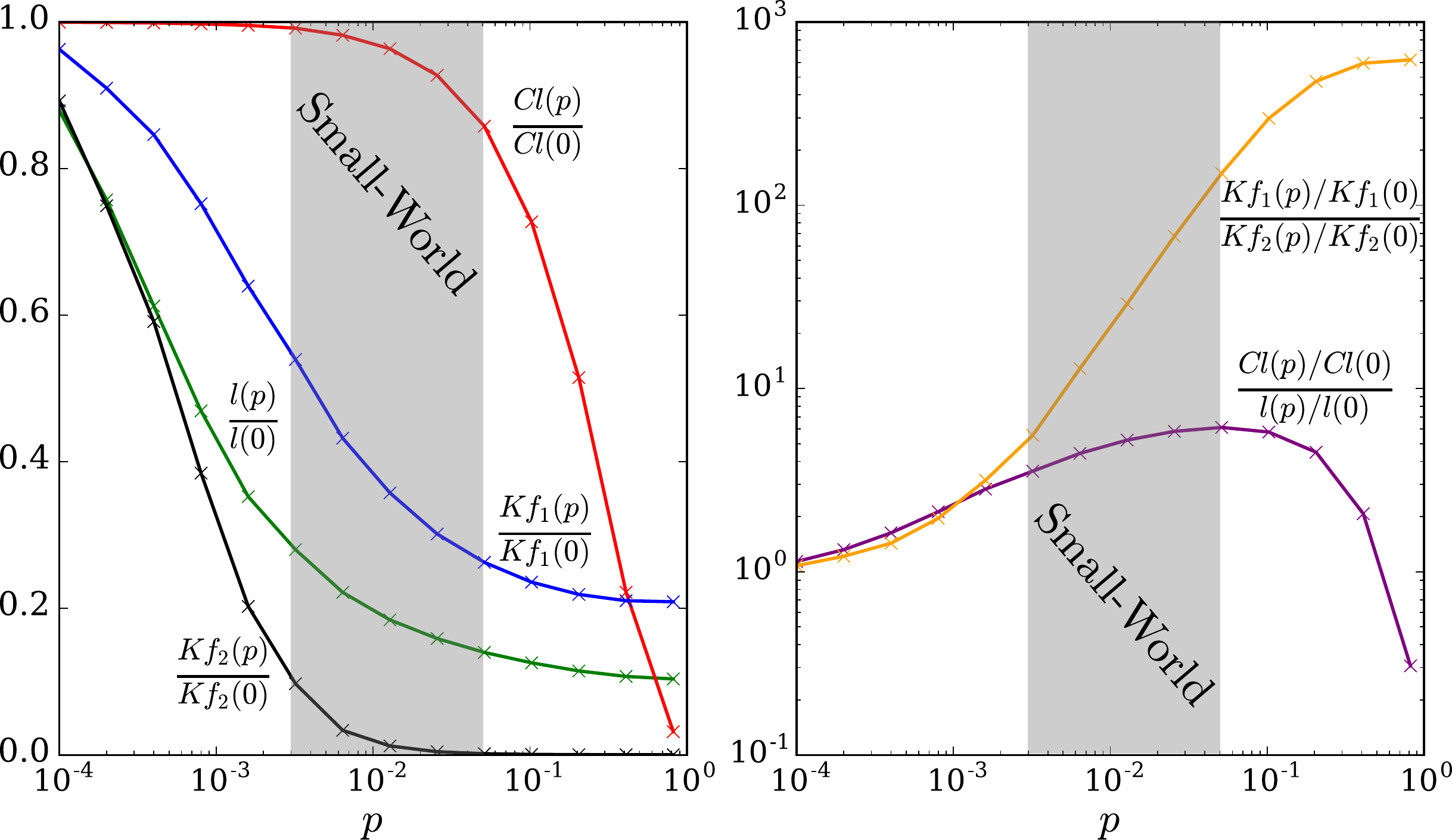}
\caption{Left panel: clustering coefficient $Cl$, geodesic centrality $l$ 
and generalized Kirchhoff indices $\Kf_1$ and  $\Kf_2$, as a function of the rewiring probability $p$
for Strogatz-Watts rewired networks~\cite{Wat98}. Each data point corresponds to an average over
$30$ realizations of randomly rewired graphs, obtained from an initial cycle graph with $n=1000$ nodes and nearest to $10^{\rm th}$- neighbor coupling, where each edge is randomly rewired with a 
probability $p$.
Right panel: ratio of the Kirchhoff indices and of clustering coefficient vs. geodesic centrality. 
Small-world network are easily identified by the steepest slope of the orange line.}\label{fig8}
\end{center}
\end{figure}

The results obtained above relate local vulnerabilities to nodal centralities and global network 
robustness
to generalized Kirchhoff indices. This connection is powerful : 
it gives a vulnerability ranking of nodes and provides robustness assessment based on 
well-defined, easily calculated network descriptors. To gain qualitative insight on 
what favors robustness in a graph, we investigate the behavior of the Kirchhoff indices
for Watts-Strogatz, randomly rewired networks. Following Ref.~\cite{Wat98}, we consider
initially regular, circular graphs where nodes are coupled to their nearest, second-nearest aso.
up to their 10$^{\rm th}$ neighbors. Each edge in the corresponding coupling network 
is then rewired with probability $p$.  Fig.~\ref{fig8} compares the standard measures of 
"nearest-neighborness" of 
geodesic centrality $l$ and clustering coefficient $Cl$ with 
the generalized Kirchhoff indices $\Kf_1$ and $\Kf_2$, as a function of $p$.

Both Kirchhoff indices drop, roughly following $l$, as $p$ is increased,
with $\Kf_2$ decreasing significantly faster than $\Kf_1$ and $l$. Traditionnally, the 
"small-world" behavior occurs around $p=0.01$, where $l$ is significantly smaller than its initial value, 
while $Cl$ has not yet changed much. In that region, 
$\Kf_1$ has been reduced to $\sim 40$\% of its initial value, while $\Kf_2$ reaches only few percents
of its initial value. 
Accordingly, small-world networks are significantly more robust to external perturbations than regular networks, particularly when considering $\mathcal{P}_1^\infty$ for long quenches. Only
a fraction of edges need to be rewired to achieve a level of robustness comparable to that of
random networks. As a side-remark, we note that the ratio of Kirchhoff 
indices provides for a clear identification of small-world networks, which correspond to 
values of $p$ where $\Kf_1(p)/\Kf_2(p)$ is fast increasing with $p$. 

\begin{figure}
\includegraphics[width=0.98\columnwidth]{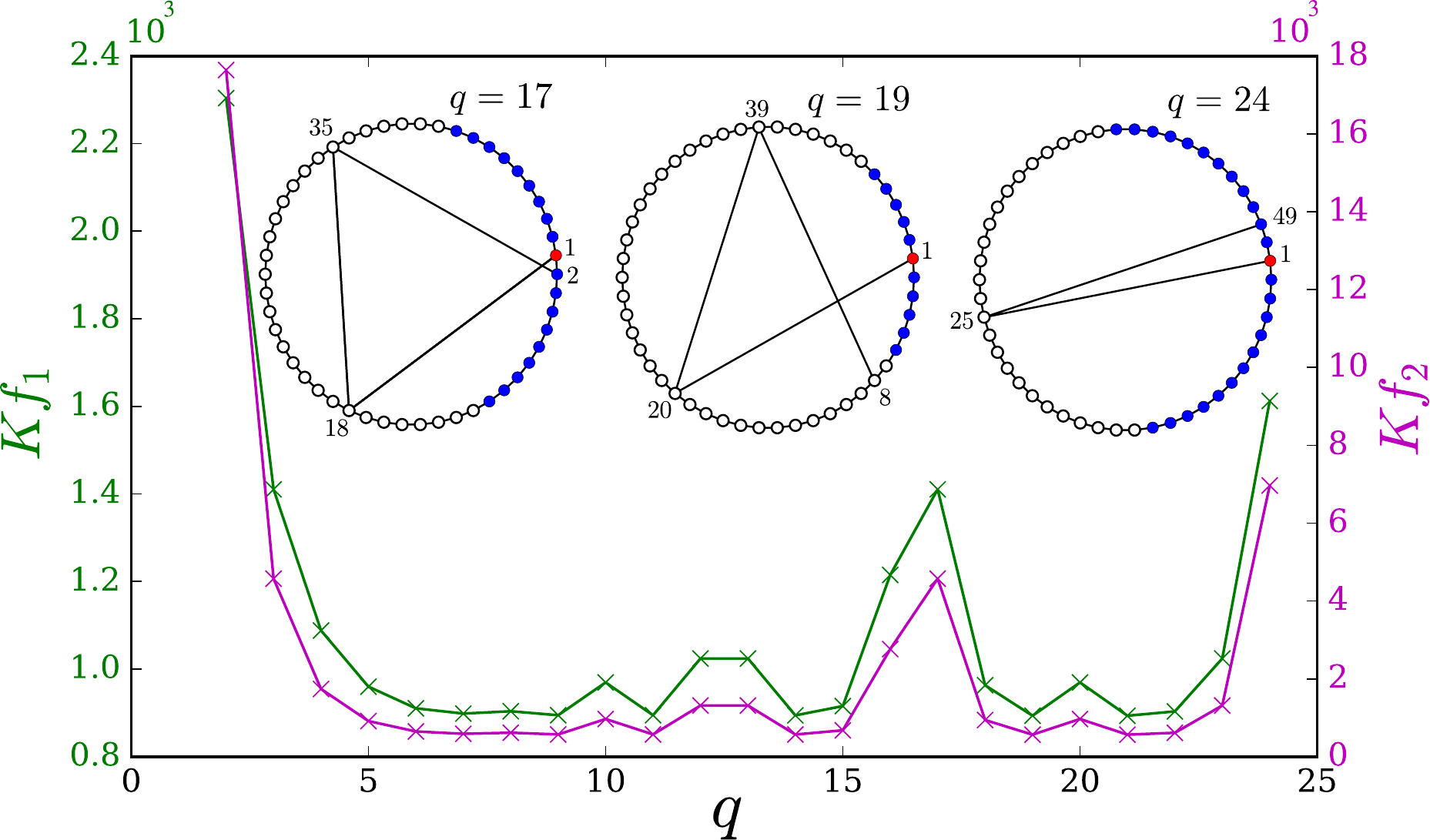}
\caption{Generalized Kirchhoff indices $\Kf_1$ (green) and $\Kf_2$ (purple)
given in Eq.~(\ref{eq:Kf_a}), for a cyclic network with $n=50$ nodes with nearest and $q^{\rm th}$- neighbor coupling. The inset sketches the model for $q=17$, $19$ and $24$ and with one path involving $q^{\rm th}$ range coupling starting from node 1 (red). The addition of the $q^{\rm th}$- neighbor coupling does not reduce geodesic distance between the reference node (red) and the set of nodes colored in blue.}\label{fig9}
\end{figure}

\subsection{Regular Networks}\label{section5}

We finally comment on regular networks. 
In such networks, all the nodes are equivalent and therefore global robustness 
is equivalent to local vulnerability, 
$\mathcal{P}_{1,2}^\infty(k)=\langle \mathcal{P}_{1,2}^\infty \rangle$, $\forall k$,
furthermore,  Kirchhoff indices can be calculated
analytically. The Laplacian matrix can be diagonalized with a Fourier transform, and 
its spectrum is given by 
\begin{eqnarray}
\lambda_\alpha=4-2\cos(k_\alpha)-\cos(k_\alpha q) \; , \; \alpha=1,...,n,
\end{eqnarray}
with $k_\alpha=2\pi(\alpha-1)/n$. Kirchhoff indices Eq.~(\ref{eq:Kfp}) are then given by,
\begin{eqnarray}\label{eq:Kf_a}
\Kf_p=n\sum_{\alpha\ge 2}[4-2\cos(k_\alpha)-\cos(k_\alpha q)]^{-p} \; .
\end{eqnarray}

Fig.~\ref{fig9} shows $\Kf_1$ and $\Kf_2$ for such regular networks with $n=50$ nodes.
When extending the coupling range $q$, Kirchhoff indices are generally decreasing, indicating 
the standard trend that longer-range couplings reduce centralities. However, 
for some values $q=10,17,24$ equal or close to integer divisors of $n$, Kirchhoff 
indices suddenly become larger. This is so, since then, paths made of few long-range 
interactions form either closed or almost closed loops (see the inset of Fig.~\ref{fig9} for $q=17,24$), 
which do not reduce the geodesic distance between many pairs of nodes, compared to long range coupling with $n/q$ not integer (e.g. $q=19$ in Fig.~\ref{fig9}). Consequently, graphs that may
appear similar, such as those with $q=17$ and $q=19$ or with $q=23$ and $q=24$ 
may exhibit Kirchhoff indices differing by factors of 2-4 or even more. This illustrates how
assessing 
global robustness is hard to do from a network's general appearance and/or 
from arguments solely based on the existence of long-range couplings.

\section{Conclusion}\label{section7}

Building up on earlier works~\cite{Bam12,Gru18,Poo17,Pag17,Sia16,Tyl18a,Tyl18b},
we have investigated the response under external perturbations of network-coupled
dynamical systems initially in a stable synchronous state.
We proposed to assess network robustness and identify nodal vulnerabilities through 
quadratic performance metrics quantifying the magnitude of the perturbation-induced 
transient excursion away from the synchronous state. 
As we reported earlier for first-order oscillators \cite{Tyl18a}, we found that the response of 
inertiaful, second-order oscillators depends on the overlap between the perturbation vector 
and the eigenmodes of the weighted Laplacian. 
In particular, the set of nodes located on the slowest eigenmode corresponding to the smallest
 eigenvalue produces the largest excursions when perturbed. 
Considering disturbances localized on a single node we found that, oscillators which,
once perturbed, induce the largest transient excursion 
are the ones with smallest resistance centralities. Extending the results of Ref.~\cite{Tyl18a} to second-order oscillators, we found that global robustness, assessed by averaging performance measures over ergodic ensembles of perturbation vectors, is also given by generalized Kirchhoff indices. A network can then be made more robust to perturbations by minimizing its average resistance distances, for
instance by introducing long-range edges. Quite remarkably, except for $\mathcal{P}_2$ and short time perturbation, asymptotic behaviors of performance measures in either limit of long or short
perturbations
do not depend on the inertia of the oscillators. 

Our findings are rather general. Together with Refs.~\cite{Tyl18a,Tyl18b}, they make it clear
that, almost regardless of the presence of inertia, and of the type of perturbation
chosen, quadratic performance measures 
are given by the generalized resistance distance-based centralities or, once averaged over
ergodic ensembles of perturbations, by the generalized Kirchhoff indices that we introduced in Section~\ref{section2}. These local and global network characteristics therefore provide 
well-defined, numerically easy to calculate robustness descriptors and local vulnerability 
indicators. 

Further studies could consider the effect of spatially correlated perturbations and go beyond the
assumption of homogeneous inertia and damping.

\section{Acknowledgments}
This work has been supported by the Swiss National Science Foundation under Grant PYAPP2\_154275.

%



\begin{thebibliography}{50}
\bibitem [{rem()}]{remark1}%
  \BibitemOpen
  \href@noop {} {\bibinfo  {journal} {By some abuse of language, we often refer
  to $P_i$ as natural frequencies}\ }\BibitemShut {NoStop}%
\end{thebibliography}
\end{document}